\documentclass[letterpaper, 10 pt, conference]{ieeeconf}  

\IEEEoverridecommandlockouts
\usepackage{cite}
\usepackage{hyperref} 
\usepackage{url}      
\usepackage{amsmath,amssymb,amsfonts}
\usepackage{algorithm}
\usepackage{algorithmic}
\usepackage{graphicx}
\usepackage{textcomp}
\usepackage{xcolor}
\usepackage{tensor}
\usepackage{caption}
\usepackage{bm}
\usepackage{float}
\usepackage{subfig}
\usepackage{booktabs}
\usepackage{multirow, multicol}
\usepackage{makecell}
\usepackage{array}
\newcolumntype{C}[1]{>{\centering}p{#1}}
\usepackage{hyperref}
\hypersetup{hidelinks,
	colorlinks=true,
	allcolors=black,
	pdfstartview=Fit,
	breaklinks=true}

\def\BibTeX{{\rm B\kern-.05em{\sc i\kern-.025em b}\kern-.08em
    T\kern-.1667em\lower.7ex\hbox{E}\kern-.125emX}}

\begin{document}

\title{D-Compress:  Detail-Preserving LiDAR Range Image Compression for Real-Time Streaming on Resource-Constrained Robots}

\author{Shengqian Wang, Chang Tu, and He Chen
\thanks{The authors are with the Department of Information Engineering, The Chinese University of Hong Kong (CUHK), Hong Kong, China. Email: \{ws021, he.chen\}@ie.cuhk.edu.hk. }
}

\maketitle

\begin{abstract}
Efficient 3D LiDAR point cloud compression (LPCC) and streaming are critical for edge server-assisted robotic systems, enabling real-time communication with compact data representations. A widely adopted approach represents LiDAR point clouds as range images, enabling the direct use of mature image and video compression codecs. However, because these codecs are designed with human visual perception in mind, they often compromise geometric details, which downgrades the performance of downstream robotic tasks such as mapping and object detection. Furthermore, rate-distortion optimization (RDO)-based rate control remains largely underexplored for range image compression (RIC) under dynamic bandwidth conditions. To address these limitations, we propose D-Compress, a new detail-preserving and fast RIC framework tailored for real-time streaming. D-Compress integrates both intra- and inter-frame prediction with an adaptive discrete wavelet transform approach for precise residual compression. Additionally, we introduce a new RDO-based rate control algorithm for RIC through new rate-distortion modeling.  Extensive evaluations on various datasets demonstrate the superiority of D-Compress, which outperforms state-of-the-art (SOTA) compression methods in both geometric accuracy and downstream task performance, particularly at compression ratios exceeding 100×, while maintaining real-time execution on resource-constrained hardware. Moreover, evaluations under dynamic bandwidth conditions validate the robustness of its rate control mechanism.


\end{abstract}

\section{INTRODUCTION} \label{sec: intro}
The advancement of edge computing enables resource-constrained robots, assisted by edge servers, to be deployed in computation-intensive LiDAR applications, such as edge-based 3D object detection~\cite{lianides20223d}, LiDAR odometry and mapping (LOAM)~\cite{qingqing2019edge}, and exploratory navigation~\cite{li2024edge}. Although edge computing provides cost efficiency and high scalability, it introduces a critical challenge: ensuring real-time transmission of sequential LiDAR point clouds under dynamic bandwidth conditions, as local robots rely on timely feedback from edge servers to support decision-making and close the control loop. This challenge necessitates efficient LiDAR point cloud compression (LPCC) and streaming that simultaneously achieve high compression ratios, high accuracy, and real-time performance on resource-limited robots.

Traditional geometry-based LiDAR point cloud compression (G-LPCC) methods~\cite{schnabel2006octree, hornung2013octomap, schwarz2018emerging} leverage tree structures to compress 3D point clouds, demonstrating notable performance. Nevertheless, their high computational complexity makes them unsuitable for real-time streaming on resource-constrained robotic platforms. Furthermore, many G-LPCC approaches miss exploiting temporal redundancy between consecutive point clouds, or they depend on precise odometry or computation-intensive relative pose estimation for inter-prediction, which further limits their practical applications.

\begin{table}
    \centering
    \caption{\textit{Performance comparison between D-Compress (Ours) with other SOTA LPCC codecs at bits per point (bpp) of 1.55, including PSNR, LiDAR mapping error, 3D object detection average precision (AP) and frames per second (FPS). All experiments are evaluated on a low-cost mini PC with Intel(R) Core(TM) i5-7260U CPU @ 2.20GHz and 16 GB of RAM), using the KITTI dataset.}}
        \begin{tabular}{c|c|c|c|c}
            \hline
            Method & \makecell{PSNR \\ (dB) $\uparrow$} & \makecell{Mapping \\ Error $\downarrow$} & \makecell{Detection \\ Precision $\uparrow$} & FPS $\uparrow$ \\
            \hline
            G-PCC~\cite{schwarz2018emerging} & 63.7 & 2.0 $\%$ & 45.5 $\%$ & 1.8 \\
            \hline
            Draco~\cite{draco} & 58.9 & 4.2 $\%$ & 33.1 $\%$ & 26.2 \\
            \hline
            H.265~\cite{sullivan2012overview} & 63.4 & 2.7 $\%$ & 45.2 $\%$ & 5.1 \\
            \hline
            JPEG2000~\cite{taubman2002jpeg2000} & 58.1 & 3.7 $\%$ & 37.7 $\%$ & 27.1 \\
            \hline
            WebP~\cite{webp} & 56.3 & 5.8 $\%$ & 30.0 $\%$ & 25.7 \\
            \hline
            RT-ST~\cite{feng2020real} & 60.7 & 3.6 $\%$ & 37.3 $\%$ & 8.8 \\
            \hline
            \textbf{D-Compress} & \textbf{67.6} & \textbf{1.4 $\%$} & \textbf{53.6 $\%$} & \textbf{25.1} \\
            \hline
            Raw data & -- & 0.6 $\%$ & 62.2 $\%$ & -- \\
            \hline
        \end{tabular}
    \label{table: SOTA performance}
    \vspace{-1.5em}
\end{table}

As an alternative data representation, range images, which are generated by projecting Cartesian LiDAR points into spherical coordinates, have gained increasing popularity. A notable advantage of range images is their ability to be efficiently compressed using well-established image/video codecs~\cite{sullivan2012overview, taubman2002jpeg2000, webp}, achieving high compression ratios and real-time performance. However, these methods were originally designed for human visual perception. For instance, popular codecs achieve high compression ratios by heavily quantizing high-frequency components, leveraging the human eye’s reduced sensitivity to fine details. Yet distortions that are imperceptible to humans can lead to significant spatial displacement in reconstructed LiDAR 3D point clouds, adversely affecting downstream tasks' performance.

Table~\ref{table: SOTA performance} highlights the critical role of range image details in preserving 3D point cloud fidelity. State-of-the-art (SOTA) codecs show notable performance degradation in both LiDAR mapping and 3D object detection when operating on compressed-reconstructed point clouds compared to their uncompressed counterparts. We note that a plane-fitting approach is proposed to compress coplanar points with high accuracy by leveraging LiDAR point clouds' geometric properties~\cite{feng2020real}. Nevertheless, its accuracy declines rapidly as compression ratios increase. Recent learning-based methods (e.g., \cite{tu2019point, zhou2022riddle}) employ neural networks to learn internal geometric features of range images. Despite their impressive compression performance, these approaches require considerable computational resources and generally support only single-frame compression or depend on future frames for inter-frame prediction, making them unsuitable for real-time streaming on resource-constrained robots.

This raises a fundamental research challenge: \textit{Can we develop a range image compression (RIC) method that matches the compression efficiency of image and video codecs while preserving the geometric fidelity of LiDAR point clouds?} To address this challenge, we introduce D-Compress, a new RIC framework designed to preserve fine-grained range image details while simultaneously achieving high compression ratios and speed. D-Compress incorporates gradient-based intra-frame prediction and LiDAR motion-based inter-frame prediction with two modes switched by a keyframe selection strategy. For detail preservation, we develop an adaptive DWT (a-DWT) approach for precise residual compression. Additionally, for real-time point cloud streaming under dynamic bandwidth conditions, effective rate control is essential to determine the optimal compression level that ensures streaming quality: Over-compression causes unnecessary information loss, while under-compression produces file sizes exceeding bandwidth constraints, inducing increased transmission delays and higher packet loss rates due to channel congestion. While rate control is effectively addressed through rate-distortion optimization (RDO) in video coding, a clear research gap remains in applying RDO-based rate control within existing RIC methods. To bridge this gap, we develop a rate-distortion model and an RDO solver tailored to RIC for robust and efficient rate control.

Our main contributions are four-fold: (1) We propose D-Compress, a new LiDAR range image compression framework for real-time streaming on resource-constrained robots, which can preserve range images' geometric details while achieving high compression ratios and computation efficiency. (2) An adaptive DWT approach is developed for range image compression, prioritizing machine-level precision over human-perceptual accuracy. (3) We propose an RDO-based rate control algorithm, supported by the first RIC-specific rate-distortion (R-D) model, that adaptively adjusts each frame’s compression to match its target bitrate. (4) Extensive experiments show that D-Compress outperforms SOTA codecs in accuracy and application-level performance at high compression ratios, while running in real time on resource-limited hardware. Rate control evaluations further confirm its robustness under varying bandwidth conditions.

\begin{figure*}
    \centering
    \includegraphics[width=6.85in]{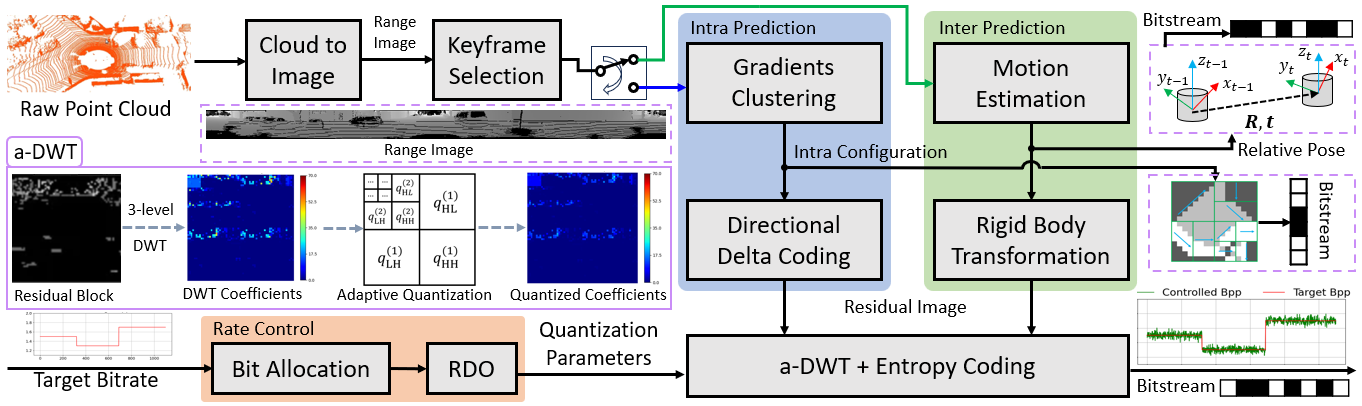}
    \caption{D-Compress overview.}
    \label{fig: system overview}
     \vspace{-1.5em}
\end{figure*}

\section{RELATED WORKS}
This section reviews existing LPCC works, categorized into geometry-based compression and range image compression. Rate control algorithms for PCC are also discussed.

\textbf{Geometry-Based Methods}: Tree structures have been extensively employed in G-LPCC methods, with the octree being the most predominant. Numerous studies~\cite{schnabel2006octree, hornung2013octomap, schwarz2018emerging, golla2015real} use octrees as foundational 3D representations, developing encoding techniques that exploit inherent spatial redundancy. Beyond octree, alternative structures like k-d tree~\cite{draco} and prediction tree~\cite{gumhold2005predictive} have also been adopted for compression. Recent deep learning approaches~\cite{huang2020octsqueeze, que2021voxelcontext, fu2022octattention} achieve notable performance gains by learning geometric priors. OctSqueeze~\cite{huang2020octsqueeze} designs a tree-structured deep conditional entropy model to predict octree symbol probabilities for efficient coding. Nevertheless, these methods exhibit high computational complexity, which limits their deployments on resource-constrained robots.

\textbf{Range Image Compression}: An alternative representation converts 3D point clouds into 2D range images via spherical or orthogonal projection. Range images allow LiDAR data to be compressed using standard image/video codecs (e.g., H.265~\cite{sullivan2012overview}, JPEG2000~\cite{taubman2002jpeg2000}, WebP~\cite{webp}). However, as discussed in Sec.~\ref{sec: intro}, the human vision-centric precision and integer-only intensity representation of these codecs are suboptimal for LiDAR data, leading to degraded performance in downstream tasks. An alternative approach is the accurate plane-fitting algorithm proposed in~\cite{feng2020real} for range image coding with inter-frame compression. However, its practical utility is limited by relatively low compression ratios and a reliance on inertial measurement unit (IMU) data. Tu et al.~\cite{tu2019point} realized high compression ratios of range images using recurrent neural network (RNN) and residual blocks. RIDDLE~\cite{zhou2022riddle} employs deep 4D context modeling to predict pixel values with entropy-coded residuals. However, these learning-based methods face real-time deployment challenges on resource-constrained robots.

\textbf{Rate Control}: Rate control has been widely researched in video coding~\cite{chang2008q, wang2013quadratic, li2014lambda}, primarily through RDO. Drawing inspiration from these approaches, various vision-based PCC (V-PCC) studies~\cite{liu2021reduced, li2020rate} develop rate control algorithms tailored to dense point clouds. For sparse LiDAR-acquired point clouds, Li et al.~\cite{li2022frame} modeled the R-D relationship using point density as a quantization parameter, a method validated across multiple datasets. Building on this model, they developed the G-LPCC rate control algorithm. Hou et al.~\cite{hou2024rate} defined the quantization parameter as the minimum distance between quantized points in predictive tree-based G-LPCC, proposing a $l$-domain rate control method. However, rate control through R-D modeling and RDO for range image-based LPCC remains underexplored.

\section{D-COMPRESS DESIGN}
\subsection{Overview}
As depicted in Fig.~\ref{fig: system overview}, D-Compress first converts point clouds into 2D range images using spherical coordinates, as described in~\cite{feng2020real}. Range images then pass through a keyframe selection module, which adaptively determines the compression mode for each range image. This adaptive selection strategy, often missing in existing LPCC methods that typically use a fixed keyframe sampling frequency, is important because assigning appropriate compression modes improves both compression accuracy and overall compression ratio. Specifically, our method follows a principle commonly used in video coding: when a large inter-prediction residual is observed, indicating weak temporal correlation with the previous frame, the current frame is treated as a keyframe and switches to intra-frame prediction.

\begin{figure}
	\centering
	  \subfloat[\label{fig: frame prediction}]{
		\includegraphics[scale=0.223]{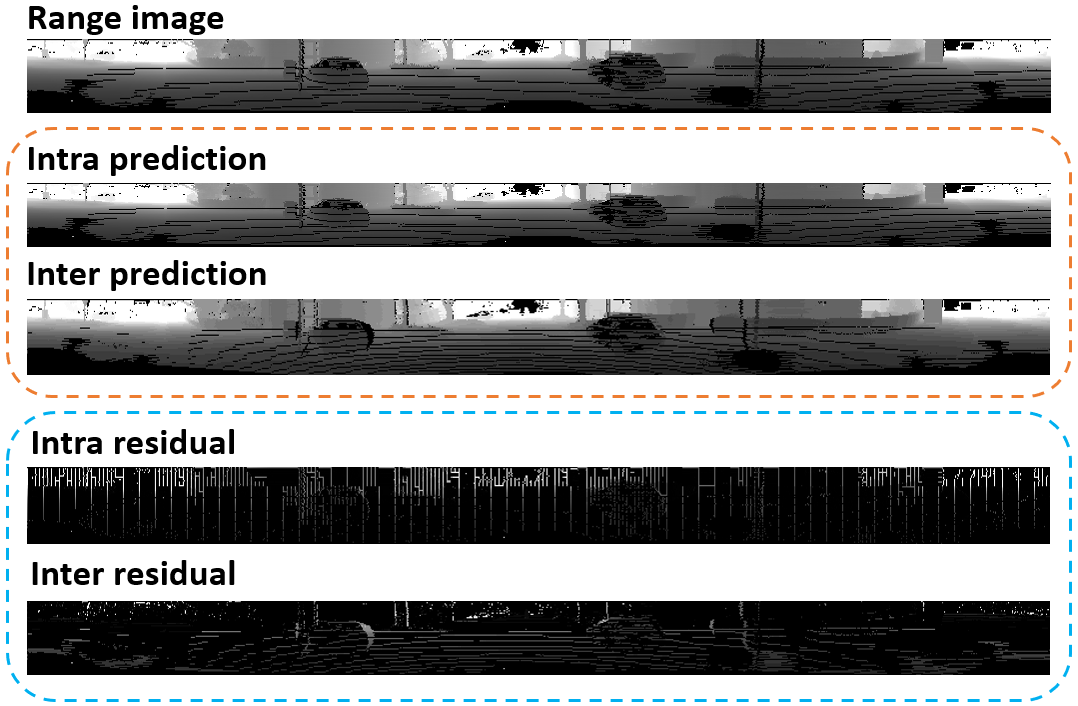}} \\

        \subfloat[\label{fig: a-DWT}]{
		\includegraphics[scale=0.223]{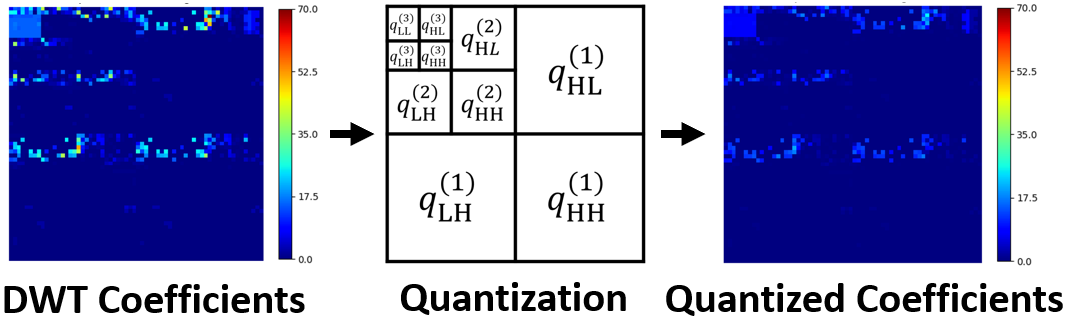}}

	\caption{Workflow of (a) Frame prediction. (b) a-DWT. }
	\label{fig: compression illustration}
	 \vspace{-2em}
\end{figure}

For intra-frame prediction, we design a fast and efficient gradient-based method, leveraging image gradients to characterize spatial redundancy within range images and employing delta coding for prediction. For inter-frame prediction, temporal redundancy between consecutive frames is exploited using a rigid body transformation derived from odometry. When odometry information is unavailable or inaccurate, D-Compress employs a fast and robust relative pose estimation method. The residual images generated by frame prediction are compressed using the proposed a-DWT approach in a detail-preserving manner. This method adaptively assigns quantization parameters to each frequency subband based on its carried information content. To operate under dynamic bandwidth constraints, D-Compress models the R-D characteristics of range image compression and develops an RDO-based rate control algorithm to determine the optimal compression levels ensuring that the compressed bitrates conform to the target values. Fig.~\ref{fig: compression illustration} depicts the intermediate results at various stages of the processing pipeline.



\subsection{Intra-Frame Prediction} \label{subsection: intra}

Spatial redundancy in range images mainly arises from planar surfaces, where neighboring pixels have similar distance values. Intra prediction in video coding~\cite{sullivan2012overview} exploits this similarity but requires exhaustive R-D cost evaluations for mode selection, causing high computational overhead on resource-limited devices. A plane fitting method~\cite{feng2020real} encodes similar pixels within macroblocks by fitting surface functions, but its row-wise approach reduces compression efficiency, especially at high compression ratios where coarse approximations degrade accuracy.

\begin{figure}
    \centering
    \includegraphics[width=3.3in]{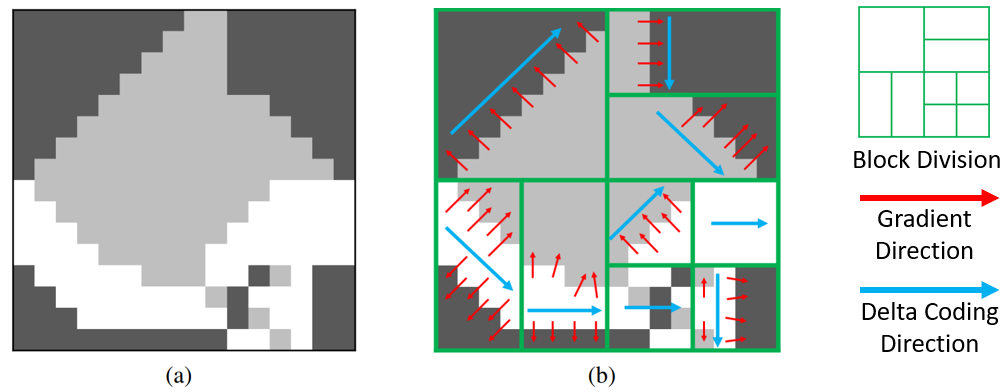}
    \caption{Illustration of gradient clustering: (a) A macroblock. (b) Block division and delta coding direction.}
    \label{fig: gradient-based intra}
     \vspace{-1.5em}
\end{figure}

To address these limitations, we propose an intra prediction method inspired by the observation that pixel gradients reveal local intensity changes. The direction orthogonal to the dominant gradient within a macroblock captures pixel similarity. Applying delta coding~\cite{mathias1994changing} along this coherence direction yields near-zero differences, enabling efficient compression. As shown in Fig.~\ref{fig: gradient-based intra}, each range image is divided into 16×16 macroblocks. We compute the gradient $\nabla I(u, v)$ for each pixel via $\nabla I(u, v) = (\partial I(u, v)/\partial u,  \partial I(u, v)/\partial v)$, then cluster gradient orientations to determine a dominant direction per block. Delta coding is performed orthogonal to this direction. If no dominant gradient is found, the block is recursively subdivided until a clear direction emerges or a minimum size of 4×4 is reached, in which case horizontal delta coding is used by default. For an input range image using intra-frame prediction, the residual range image is obtained after this delta coding process.

\subsection{Inter-Frame Prediction}

Inter-frame prediction reduces temporal redundancy by leveraging correlations between consecutive frames. While motion estimation and compensation methods from video coding~\cite{sullivan2012overview} are applicable to range image sequences, they introduce high computational cost and suboptimal compression in LPCC due to the overhead of motion vector matrices. Instead, rigid body transformations provide a more efficient solution: each point $\mathbf{p} = [x, y, z]^T$ from the previous frame is projected to the current frame using $\mathbf{p}' = \mathbf{R}\mathbf{p} + \mathbf{t}$, where $\mathbf{R}$ and $\mathbf{t}$ represent the relative rotation and translation. This approach compactly models motion using just 12 floating-point values. If available, IMU data can directly provide these parameters. When LiDAR odometry is unreliable or missing, we use the ICP algorithm~\cite{zhang2021iterative} for motion estimation. Although standard ICP is computationally intensive and sensitive to outliers, we mitigate this by using Link3D~\cite{cui2024link3d} to extract a sparse set of matched 3D feature points $\mathcal{MP} = {(\mathbf{p}_k, \mathbf{q}_k)}_n$ between frames. These edge-based correspondences reduce computational load by avoiding dense point cloud operations and improve robustness by limiting the impact of outliers. The obtained or estimated values of $\mathbf{R}$ and $\mathbf{t}$ are subsequently quantized for efficient transmission.

With $\mathbf{R}$ and $\mathbf{t}$, the predicted 3D point cloud can be generated by transforming the previous frame's point cloud. This predicted 3D cloud is converted to a 2D predicted range image. By computing the difference between the predicted and current range images, the residual range image for inter-frame prediction is obtained.

\subsection{Residual Range Image Compression using a-DWT}
To effectively compress the residual range images produced by intra- or inter- frame prediction, we propose an a-DWT approach designed for detail-preserving compression. In contrast to conventional image and video codecs optimized for human perception, which tend to heavily quantize high-frequency components (i.e., image details), a-DWT adaptively allocates quantization precision to high-frequency subbands based on their energy proportion relative to the low-frequency subband. This is particularly important for residual range images, which typically exhibit stronger high-frequency content than natural images. We employ DWT rather than commonly-used discrete cosine transformation (DCT) because DWT provides better representation of both spatial and frequency features of residual images. In a-DWT, each residual image is divided into blocks (64x64 pixels) processed by 3-level Haar DWT, balancing frequency resolution and computational efficiency.


In the 3-level DWT, quantization steps are assigned sequentially from level 3 to level 1, as illustrated in Fig.\ref{fig: compression illustration}(b). Each level contains four subbands, denoted by $sub = \{LL, HL, LH, HH\}$. Let the quantization parameters for level-$k$ subbands be denoted as $q_{sub}^{(k)}$. Starting with the level 3 subbands, the low-frequency subband $LL^{(3)}$ contains the richest structural information of the residual range image. Its quantization step $q_{LL}^{(3)}$ serves as the reference for the other three high-frequency subbands within that level and is predetermined by the rate control algorithm introduced later in Sec.~\ref{subsection: rate control}. Given $q_{LL}^{(3)}$, the quantization step of the $HH^{(3)}$ subband, comprising both horizontal and vertical high-frequency components, is computed by
\begin{equation} \label{eq: qhh}
    q_{HH}^{(3)} = \alpha \log_2 (E_{LL}^{(3)}/E_{HH}^{(3)} + 1) \cdot q_{LL}^{(3)},
\end{equation}
where $E_{LL}^{(3)}/E_{HH}^{(3)}$ represents the energy ratio between the $LL^{(3)}$ and $HH^{(3)}$ subbands, and $\alpha$ is a positive factor, empirically set to be 0.53 in this paper. Eq.~(\ref{eq: qhh}) ensures higher quantization precision for the $HH^{(3)}$ subband when it exhibits greater energy, indicating the presence of richer information. The logarithmic function is empirically adopted here to establish a robust monotonic relationship between $q_{HH}^{(3)}$ and $E_{HH}^{(3)}$. For the $LH^{(3)}$ and $HL^{(3)}$ subbands, which contain both high- and low-frequency components, their quantization steps are formulated as the weighted average of $q_{LL}^{(3)}$ and $q_{HH}^{(3)}$ as follows:
\begin{equation} \label{eq: qhl and qlh}
    q_{IJ}^{(3)} = \frac{E_{IJ}^{(3)}}{E_{HL}^{(3)} + E_{LH}^{(3)}} q_{LL}^{(3)} + \left ( 1-\frac{E_{IJ}^{(3)}}{E_{HL}^{(3)} + E_{LH}^{(3)}}\right )  q_{HH}^{(3)},
\end{equation}
where $IJ \in \{HL, LH\}$. Eq.~(\ref{eq: qhl and qlh}) assigns finer quantization to subbands richer in high-frequency content.

Proceeding to level 2, since its $LL^{(2)}$ subband consists of the four subbands of level 3, $q_{LL}^{(2)}$ is computed as a weighted average of their quantization steps, where the weights are proportional to the respective energy contributions of each subband. That is,
\begin{equation}
    q_{LL}^{(2)} = \sum_{IJ \in \{LL, HL, LH, HH  \}} q_{IJ}^{(3)}E_{IJ}^{(3)}/E_{sum}^{(3)},
\end{equation}
where $E_{\text{sum}}^{(3)}$ denotes the total energy of all subbands at level 3, which also corresponds to the energy of the $LL^{(2)}$ subband. Subsequently, the quantization steps for the remaining level-2 subbands can be determined by following the principles outlined in Eqs.~(\ref{eq: qhh})–(\ref{eq: qhl and qlh}). The quantization parameters for level 1 can then be derived iteratively by applying the same process. Quantized DWT coefficients finally undergo entropy coding for further compression.

\subsection{Rate Control} \label{subsection: rate control}
Real-time range image streaming requires precise bitrate regulation to match target bitrates determined by available bandwidth. Although rate control has been extensively studied in video coding through RDO, and such techniques have been applied to V-PCC for dense point clouds, their underlying R-D models are not fully compatible with LiDAR range images. As such, 
the resulting rate control algorithms and associated parameter updating strategies remain suboptimal for LiDAR range image compression. To bridge this gap, we, for the first time, model the range image-specific R-D characteristics and experimentally validate them using extensive point cloud data from multiple datasets.

\begin{figure}
	\centering
	\subfloat[\label{fig: D-QP curve}]{
    \includegraphics[scale=0.115]{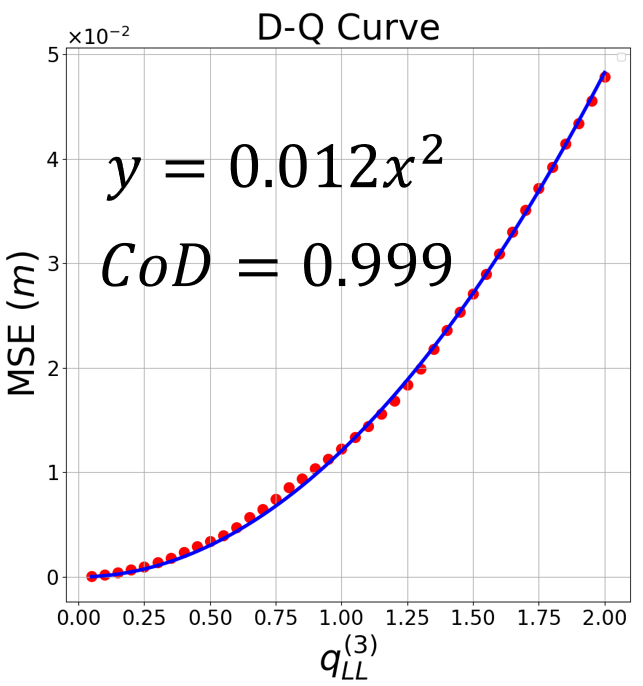}}
    \subfloat[\label{fig: R-QP curve}]{
    \includegraphics[scale=0.103]{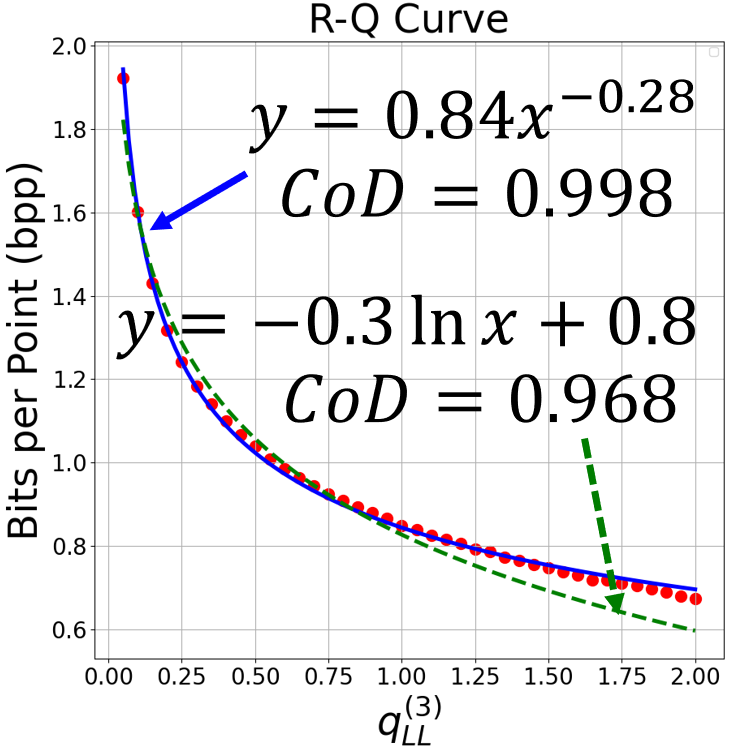}}
    \subfloat[\label{fig: lambda-QP curve}]{
    \includegraphics[scale=0.103]{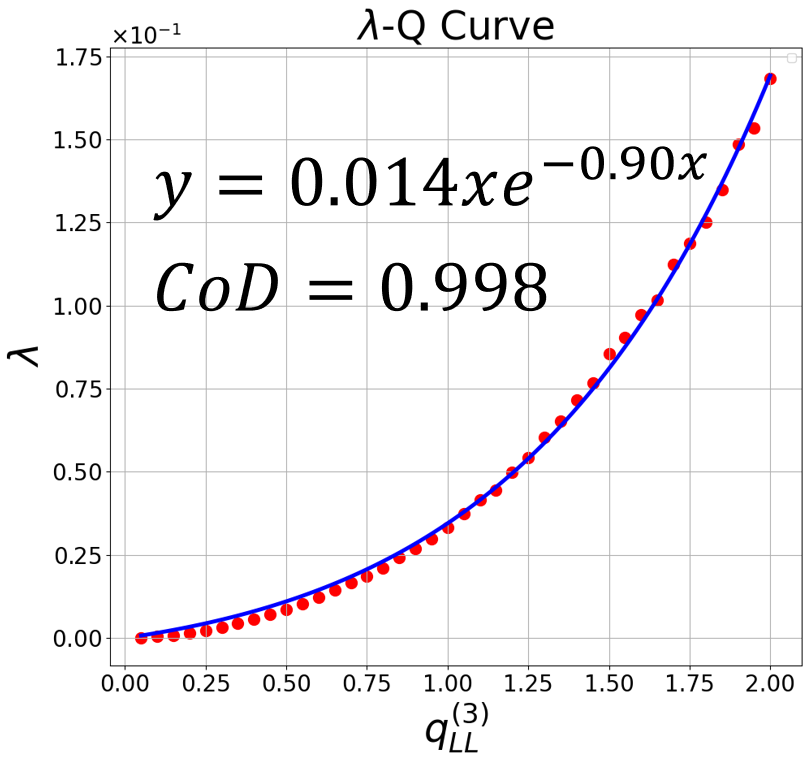}} \\
    \vspace{-1.0em}

    \subfloat[\label{fig: a_D}]{
    \includegraphics[scale=0.109]{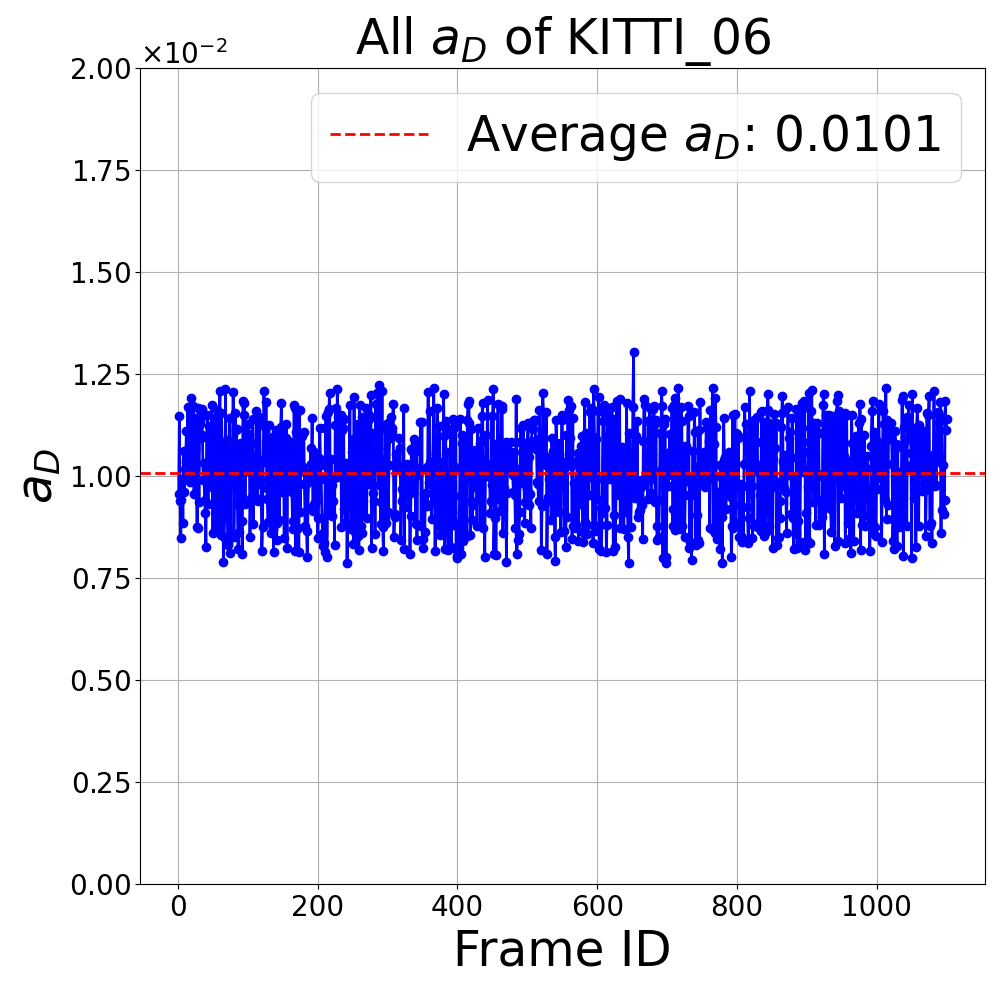}}
    \subfloat[\label{fig: a_R}]{
    \includegraphics[scale=0.109]{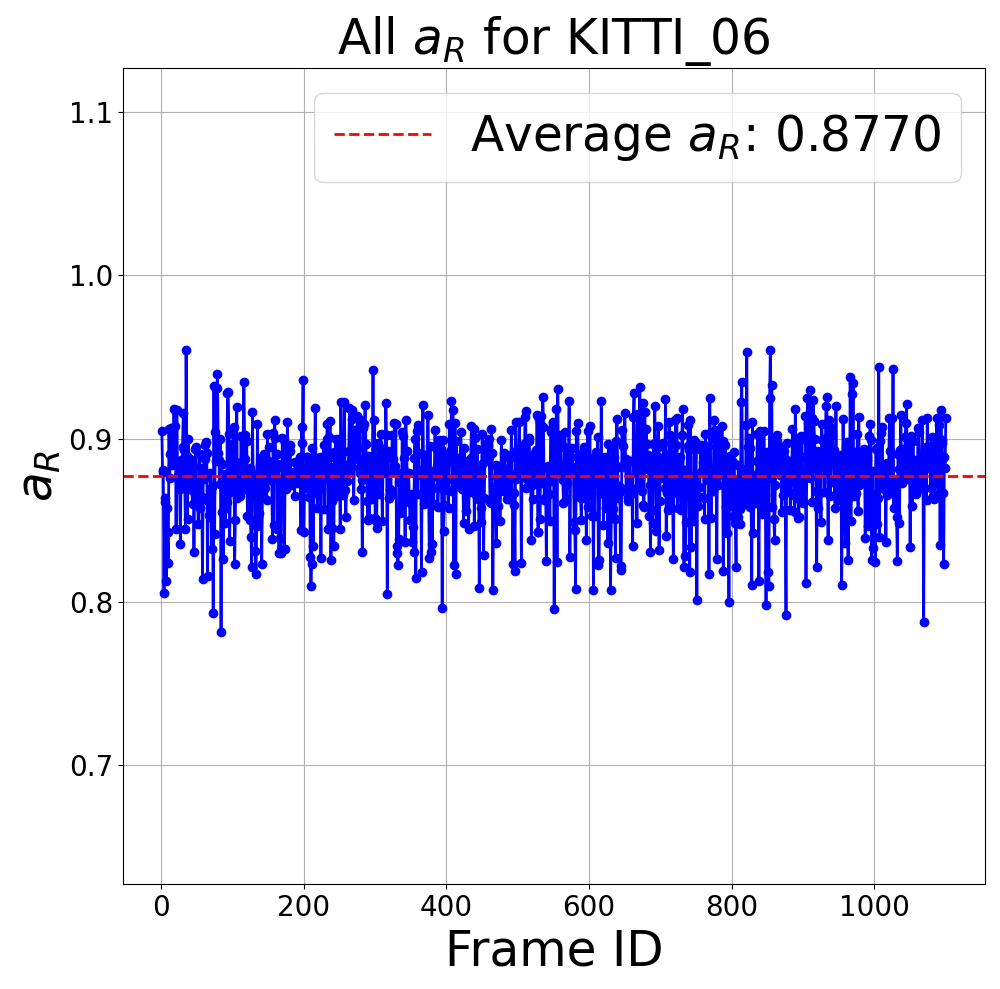}}
    \subfloat[\label{fig: b_R}]{
    \includegraphics[scale=0.109]{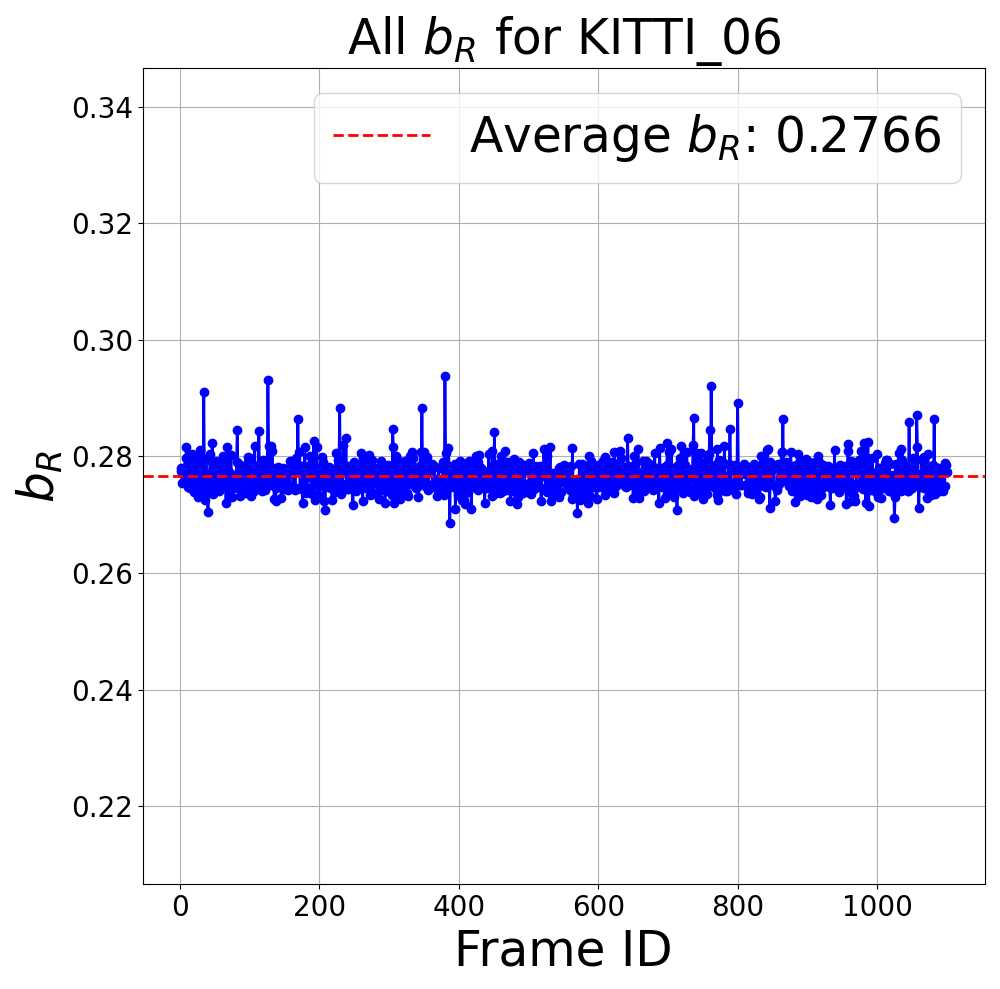}}

	\caption{R-D modeling validation by curve fitting: (a) D-Q curve. (b) R-Q curve. (c) $\lambda$-Q curve. (d)-(f) present the values of fitted $a_D$, $a_R$ and $b_R$ for all point clouds in  $\mathrm{KITTI_{06}}$ sequence, confirming the fidelity of our model fitting.}
	\label{fig: R-D modeling}
	 \vspace{-0.5em}
\end{figure}

Building on the new R-D model, we develop a new RDO-based rate control algorithm tailored for RIC. Specifically, the RDO objective seeks compression parameters that minimize compression distortion $D$ while ensuring bitrate $R$ remains below target, formulated as follows:
\begin{equation} \label{eq: rdo}
    Q^* = \mathop{\arg\min}\limits_{Q} \ J(Q) = \mathop{\arg\min}\limits_{Q} \ D(Q) + \lambda R(Q),
\end{equation}
where $\lambda = -\partial D/\partial R$ is the slope of R-D curve, and $Q$ here is the quantization parameter $q_{LL}^{(3)}$. The distortion $D$ and bitrate $R$ denote mean square error (MSE) and bits per point (bpp), respectively. For a non-zero DWT coefficient $c \neq 0$, its quantization error $e = c - {\rm{round}}(c/Q)\cdot Q$ is uniformly distributed within the range $[-Q/2, Q/2]$. Consequently, the D-Q relationship can be modeled as a quadratic function $D = \rho\int_{-Q/2}^{Q/2}e^2/Q \ de \triangleq a_D Q^2$, where $\rho$ is the proportion of non-zero coefficients. In video coding, the R-D model typically follows either an exponential function $D = ae^{-bR}$ or a hyperbolic function $D = aR^{-b}$, which indicates two candidate functions for the R-Q relationship: $R = a_R Q^{-b_R}$ or $R = -a_R\ln{Q} + b_R$. We validate the above D-Q and R-Q models on three popular LiDAR point cloud datasets: KITTI~\cite{geiger2013vision}, nuScenes~\cite{caesar2020nuscenes} and Waymo Perception~\cite{sun2020scalability}. As shown in Fig.~\ref{fig: R-D modeling} (a), (b), (d)-(f) and Table.~\ref{table: R-D curve}, the D-Q curve fits well with the quadratic function with high coefficient of determination (CoD)~\cite{di2008coefficient}, while the R-Q curve aligns better with the hyperbolic model. 
Meanwhile, for all point clouds within the same dataset, the parameters of their R-D models exhibit tightly bounded variation around certain means, with these means being listed in Table.~\ref{table: R-D curve}.

\begin{table}
    \centering
    \caption{Parameters of D-Q Models and R-Q Models}
        \begin{tabular}{c|cc|ccc}
            \hline
            \multirow{2}{*}{Dataset} & \multicolumn{2}{|c}{Quadratic model} & \multicolumn{3}{|c}{Hyperbolic model} \\
            \cline{2-6}
            & $a_D$ & CoD & $a_R$ & $b_R$ & CoD \\
            \hline
            KITTI & 0.0100 & 1.00 & 0.86 & 0.277 & 0.99 \\
            \hline
            nuScenes & 0.0109 & 0.99 & 1.82 & 0.234 & 0.99 \\
            \hline
            Waymo Perception & 0.0101 & 1.00 & 0.72 & 0.243 & 0.99 \\
            \hline
        \end{tabular}
    \label{table: R-D curve}
    \vspace{-1.5em}
\end{table}

\textit{1) Bit Allocation: } Recall that residual images undergo block-wise compression, as established in previous subsections. Given the target bits for the current frame $B_{fTar}$, we allocate bits to each block by
\begin{equation}
    B_{curTar} = \omega_{cur}B_{rem} \triangleq \frac{E_{cur}}{\sum_{notEncoded}E_{block}} B_{rem},
\end{equation}
where $B_{rem}$ denotes the remaining bits after encoding previous blocks, and $\omega_{cur}$ is the weight of the current block's energy among unencoded blocks. For the first block, $B_{rem} = B_{fTar}$. Upon completing each block's compression, $B_{rem}$ is decreased by the actual consumed bitrate.

\textit{2) Rate Control: } Based on prior R-D models, we have $\partial^2J/\partial^2Q = 2a_D+\hat{\lambda}a_Rb_R^2\exp(-b_RQ) > 0$. As such, the optimal Q for Eq.~(\ref{eq: rdo}) occurs at $\partial J/\partial Q^* = 0$, which is
\begin{equation} \label{eq: lambda-Q curve}
    \hat{Q}^* = \mathop{\arg}\limits_{Q} \ \{\hat{\lambda} - \alpha Q \exp(\beta Q) = 0\},
\end{equation}
where $\alpha$ and $\beta$ are defined to simplify the notation. Validation of the above $\lambda-Q^*$ model across datasets aligns well with Eq.~(\ref{eq: lambda-Q curve}). Specifically, we have validated the above $\lambda-Q^*$ model on the abovementioned three datasets, and the fitting results align well with Eq.~(\ref{eq: rdo}), as shown in Fig.~\ref{fig: R-D modeling} (c). Given the target bitrate $R_{curTar}$ of the current block derived from $B_{curTar}$, the estimated $\hat{\lambda}$ can be derived using a method similar to that in~\cite{li2014lambda}. However, the bounded factors $\alpha$ and $\beta$ vary across residual images for each block due to differences in content, and therefore need to be updated accordingly. In our implementation, $\alpha$ and $\beta$ are initialized to the average values of the tested three datasets, i.e., 0.014 and 0.91, respectively. Using the real bitrate and actual $\lambda$ after encoding, we can calculate the actual $Q^*_{a}$ and thereby update $\alpha$ and $\beta$ for the next frame's corresponding block via adaptive least mean squares~\cite{di2016adaptive} as follows:
\vspace{-0.5em}
\begin{equation}
\begin{aligned}
    \alpha_{t+1} &= \alpha_t + \delta_{\alpha}\alpha_t Q^*_{a}(Q^*_{a} - \hat{Q}^*)/(\beta_t Q^*_{a} + 1), \\
    \beta_{t+1} &= \beta_t + \delta_{\beta}{Q^*_{a}}^2(Q^*_{a} - \hat{Q}^*)/(\beta_t Q^*_{a} + 1),
\end{aligned}
\end{equation}
where we empirically set $\delta_{\alpha}= 0.4$ and $\delta_{\beta}= 0.3$.

\section{EVALUATION} \label{sec: evaluation}
This section presents a comprehensive experimental evaluation validating D-Compress's performance against mainstream compression methods, assessing three aspects: compression performance, downstream application efficacy and rate control performance. All experiments were conducted on a mini PC equipped with an Intel(R) Core(TM) i5-7260U CPU @ 2.20GHz and 16 GB of RAM, running Ubuntu 20.04. The KITTI dataset is used~\cite{geiger2013vision}. The baseline LPCC methods include: MPEG'S G-PCC codec~\cite{schwarz2018emerging}; range image compression via JPEG~\cite{wallace1991jpeg}, JPEG2000~\cite{taubman2002jpeg2000}, WebP~\cite{webp} and H.265~\cite{sullivan2012overview}; Google's Draco~\cite{draco}; and Feng's RT-ST~\cite{feng2020real}.

\begin{figure}
    \centering
    \includegraphics[width=3.3in]{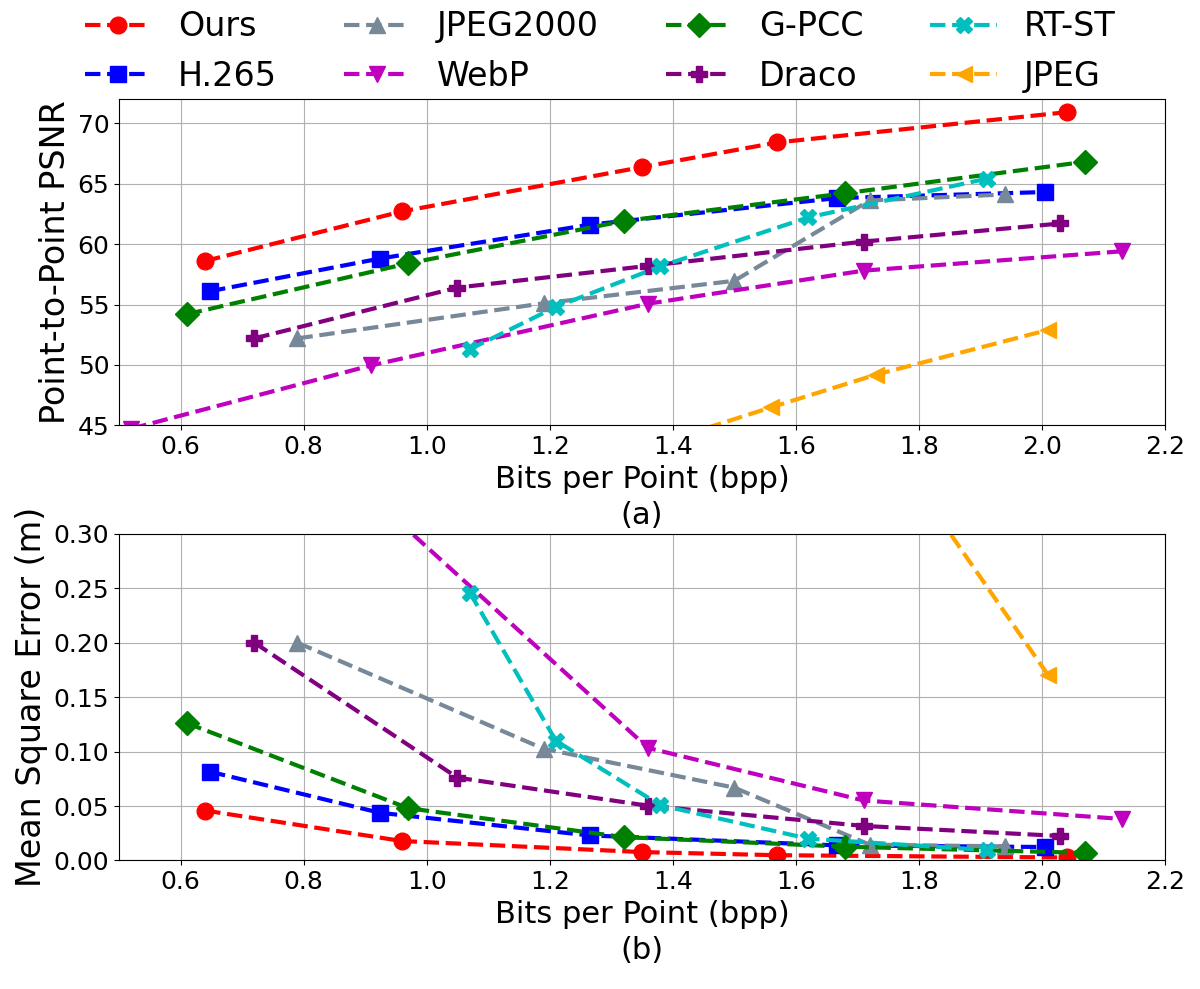}
    \caption{Compression metrics under varying bpps: (a) Point-to-point PSNR. (b) Mean square error.}
    \label{fig: compression results}
     \vspace{-1em}
\end{figure}

\begin{figure}
    \centering
    \includegraphics[width=3in]{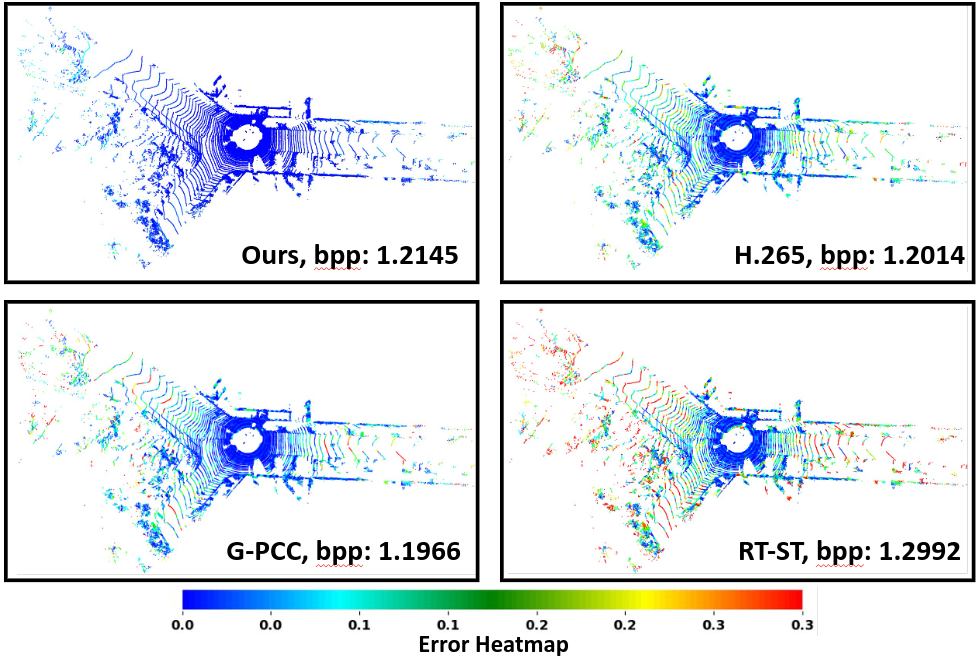}
    \caption{Visuailzed compression error of D-Compress and top-accurate baseline methods H.265, G-PCC and RT-ST. The heatmap unit is meter.}
    \label{fig: error heatmap}
     \vspace{-1.5em}
\end{figure}

\subsection{Compression Performance}

Compression performance is evaluated by point-to-point mean square error (MSE) and peak signal-to-noise ratio (PSNR) at different bpps, as well as encoding/decoding latencies. The PSNR is derived from the MSE between the original range image $P$ and decoded range image $\hat{P}$ by $\mathrm{PSNR} = 10\log_{10}(d^2_m/\mathrm{MSE}(P, \hat{P}))$, where $d_m$ is the peak value of the input point clouds.

Fig.~\ref{fig: compression results} illustrates the compression performance of the D-Compress and state-of-the-art codecs. Leveraging high-accuracy compression, D-Compress surpasses all baseline codecs in terms of PSNR and MSE across all tested bpps. Specifically, D-Compress consistently achieves the highest PSNR and the lowest MSE at high compression ratios (i.e., bpp $<$ 2.2), whereas JPEG2000 and RT-ST show rapidly degrading accuracy as the compression ratio increases. Although H.265 and MPEG's G-PCC achieve high compression accuracy among baseline methods, their encoding and decoding latencies on resource-constrained devices (see Table~\ref{table: latency tests}) fall short of meeting LiDAR's real-time requirements, typically achieving compression rates below 10 fps. Conversely, while WebP, Draco and JPEG are computationally efficient, they offer insufficient compression quality. The compression errors of D-Compress, along with those of top-performing baselines, are visualized in Fig.~\ref{fig: error heatmap}. We can see from Fig.~\ref{fig: error heatmap} that D-Compress demonstrates strong capability in preserving fine point cloud details with high fidelity.

\begin{figure}
    \centering
    \includegraphics[width=3.25in]{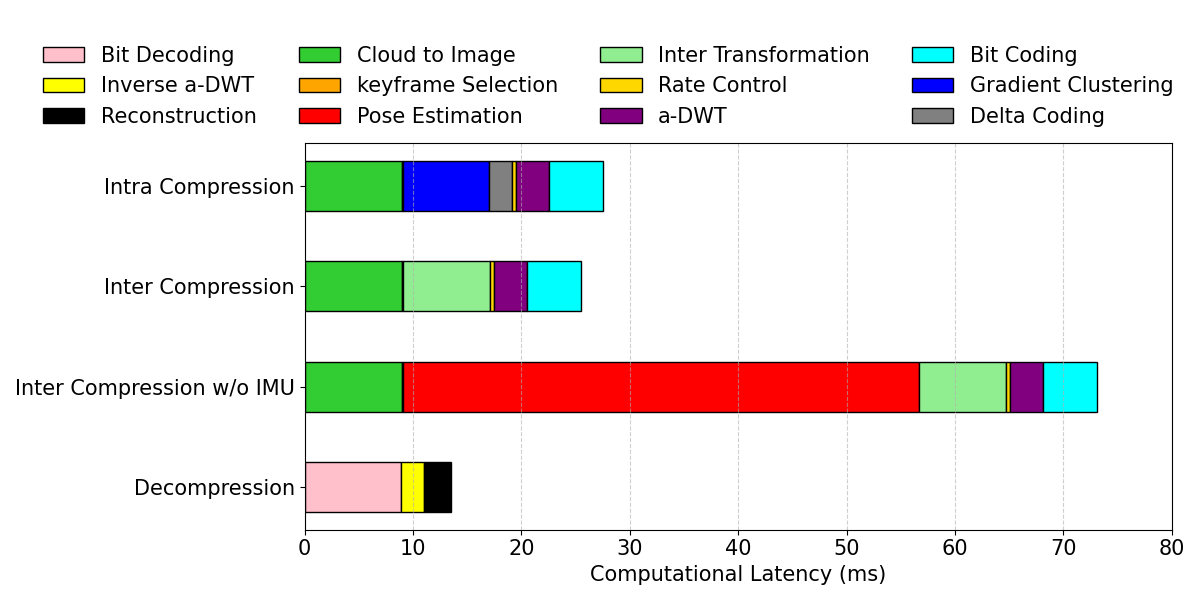}
    \caption{Computational latency analysis of D-Compress.}
    \label{fig: latency test}
     \vspace{-1.5em}
\end{figure}

In addition to achieving high compression ratios and accuracy, D-Compress meets real-time streaming requirements on edge devices, typically with runtimes under 100 ms, as shown in Table~\ref{table: latency tests} and Fig.~\ref{fig: latency test}. Notably, when integrated with IMU data, D-Compress supports high-frequency LiDAR systems (exceeding 20 fps), highlighting its potential for safety-critical applications such as high-speed drones.

\begin{table}
    \centering
    \caption{Computational latency summary}
        \begin{tabular}{c|c|c|c|c}
            \hline
            Method  & \makecell{Need \\ IMU} & \makecell{Encoding \\ Time (ms)} & \makecell{Decoding \\ Time (ms)} & \makecell{Total \\ Runtime (ms)} \\
            \hline
            WebP & $\times$ & 34.9 & 4.0 & 38.9  \\
            \hline
            x265 & $\times$ & 127.3 & 69.8 & 197.1  \\
            \hline
            RT-ST & \checkmark & 88.2 & 22.4 & 113.6  \\
            \hline
            JPEG & $\times$ & 19.4 & 0.07 & 19.4  \\
            \hline
            JPEG2000 & $\times$ & 30.9 & 6.0 & 36.9 \\
            \hline
            Draco & $\times$ & 33.6 & 4.5 & 38.1  \\
            \hline
            G-PCC & $\times$ & 359.2 & 189.4 & 548.6  \\
            \hline
            Ours with IMU & \checkmark & 26.4 & 13.5 & 39.9  \\
            \hline
            Ours w/o IMU & $\times$ & 65.9 & 13.5 & 79.4  \\
            \hline
        \end{tabular}
    \label{table: latency tests}
    \vspace{-0.5em}
\end{table}

\subsection{Downstream Application Efficacy}
We chose two representative applications for evaluation: LiDAR odometry and mapping (LOAM) and 3D object detection.

\begin{figure}
    \centering
    \includegraphics[width=3.25in]{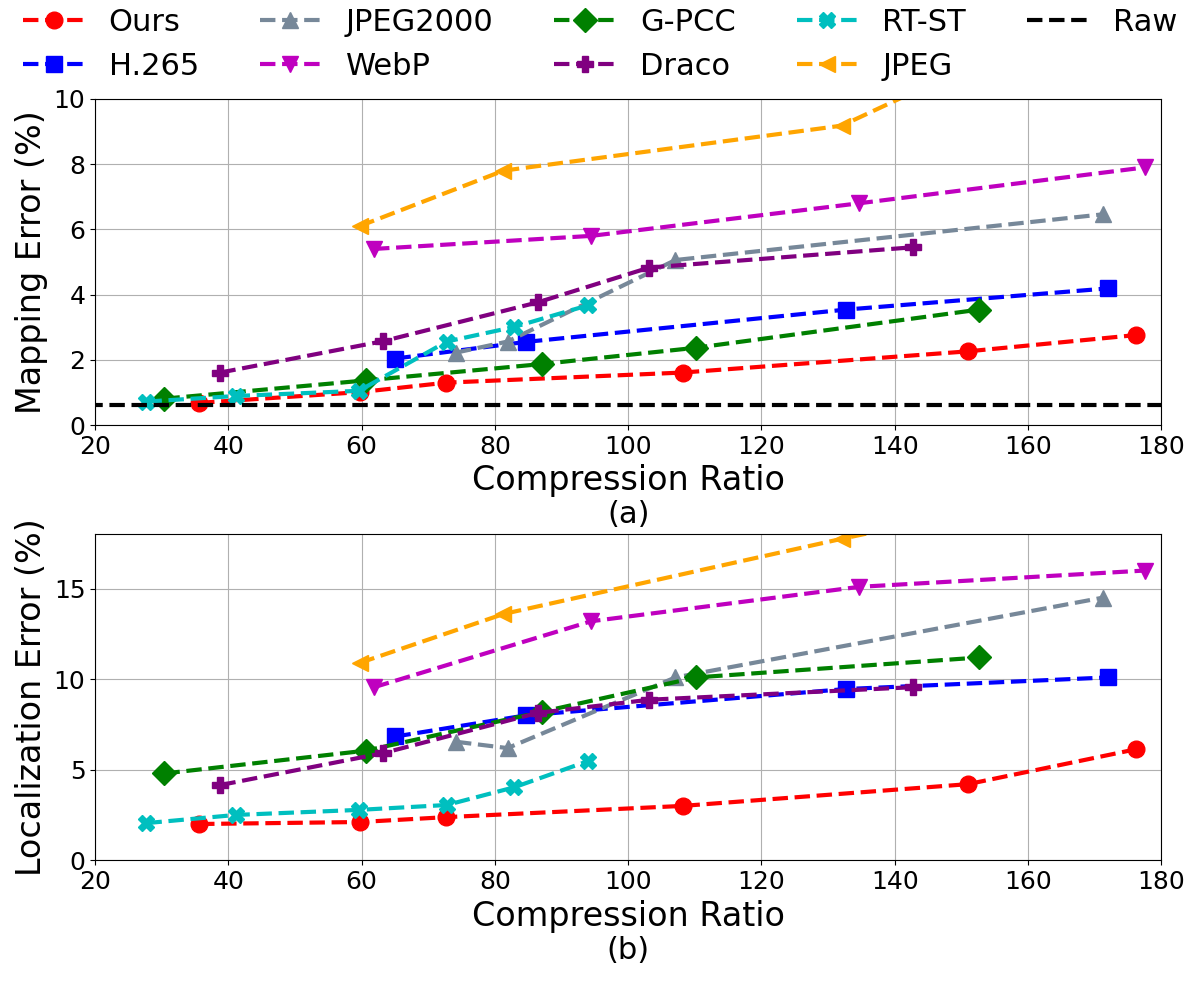}
    \caption{Translational error of (a) Mapping. (b) Localization.}
    \label{fig: loam results}
     \vspace{-1.5em}
\end{figure}

\textit{1) LOAM:} Using the A-LOAM implementation~\cite{aloam}, we process point clouds from the KITTI odometry dataset by each codec to quantify translational mapping and localization errors. As shown in Fig.~\ref{fig: loam results} (a), D-Compress maintains superior mapping accuracy with only 2.7$\%$ error at 176.3× compression ratio. In contrast, top-performing baseline methods MPEG's G-PCC and H.265 achieve merely $\sim$120× and $\sim$90× compression ratios at equivalent error levels. Localization error captures the impact of encoding and decoding latency. To align estimated positions with the ground-truth trajectory, we record timestamps marking the completion of each point cloud's codec and A-LOAM processing. Fig.~\ref{fig: loam results} (b) confirms D-Compress's consistently lowest localization error, while G-PCC and H.265 exhibit degraded accuracy due to high computational latencies. RT-ST sustains low localization error at moderate compression ratios, but experiences sharp degradation beyond 72.8× compression ratio.

\begin{figure}
    \centering
    \includegraphics[width=3.3in]{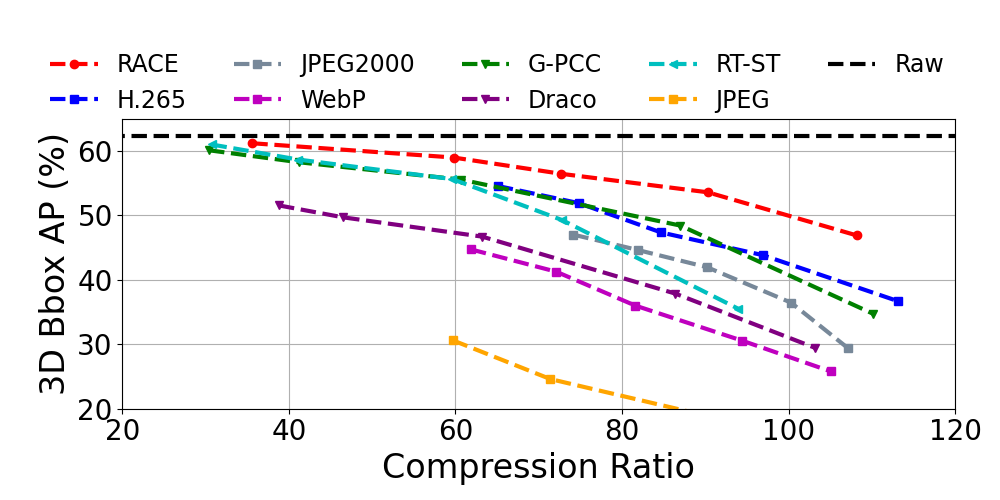}
    \caption{3D Object detection accuracy.}
    \label{fig: obj detection}
     \vspace{-1.5em}
\end{figure}

\textit{2) 3D Object Detection:} We use PointPillars~\cite{lang2019pointpillars} for 3D object detection on the KITTI detection dataset, evaluating the overall 3D bounding box average precision (3D Bbox AP). Fig.~\ref{fig: obj detection} demonstrates the detection results of different methods, with raw point clouds having 62.3$\%$ AP. D-Compress achieves the peak performance, 61.2$\%$ AP at 35.6x compression, and maintains a 48.9 $\%$ AP at 108.2x compression. Although RT-ST and G-PCC reach 61.0$\%$ and 60.1$\%$ AP respectively at $\sim$30× compression, their accuracy decline rapidly as compression ratio increases, with only 34.7 $\%$ AP at 94.0x compression and 33.9 $\%$ AP at 110.2x compression.

\subsection{Rate Control Performance}
We evaluate D-Compress's rate control performance on multiple datasets using bitrate error (BE) under dynamically constrained target bitrates. The average BE is computed as $e_R = \frac{1}{N}\sum_{i=1}^N |R_i^{real} - R_i^{target}|/R_i^{target}$,
where $R_i^{target}$ and $R_i^{real}$ denote target and real bitrates of frame $i$ and $N$ is the number of frames. We compare D-Compress with the rate control method of G-LPCC proposed by Hou et al.~\cite{hou2024rate} and the $\lambda$-domain rate control algorithm developed by Li et al.~\cite{li2014lambda}. The target bitrates for the KITTI, NuScenes, and Waymo datasets are configured as follows: 1.5 bpp, 1.5 bpp, and 2.0 bpp (first 30$\%$ frames); 1.3 bpp, 2.5 bpp, and 1.5 bpp (subsequent 40$\%$ frames); and 1.7 bpp, 2.0 bpp, and 1.0 bpp (final 30$\%$ frames), respectively.

\begin{figure}
    \centering
    \includegraphics[width=3.25in]{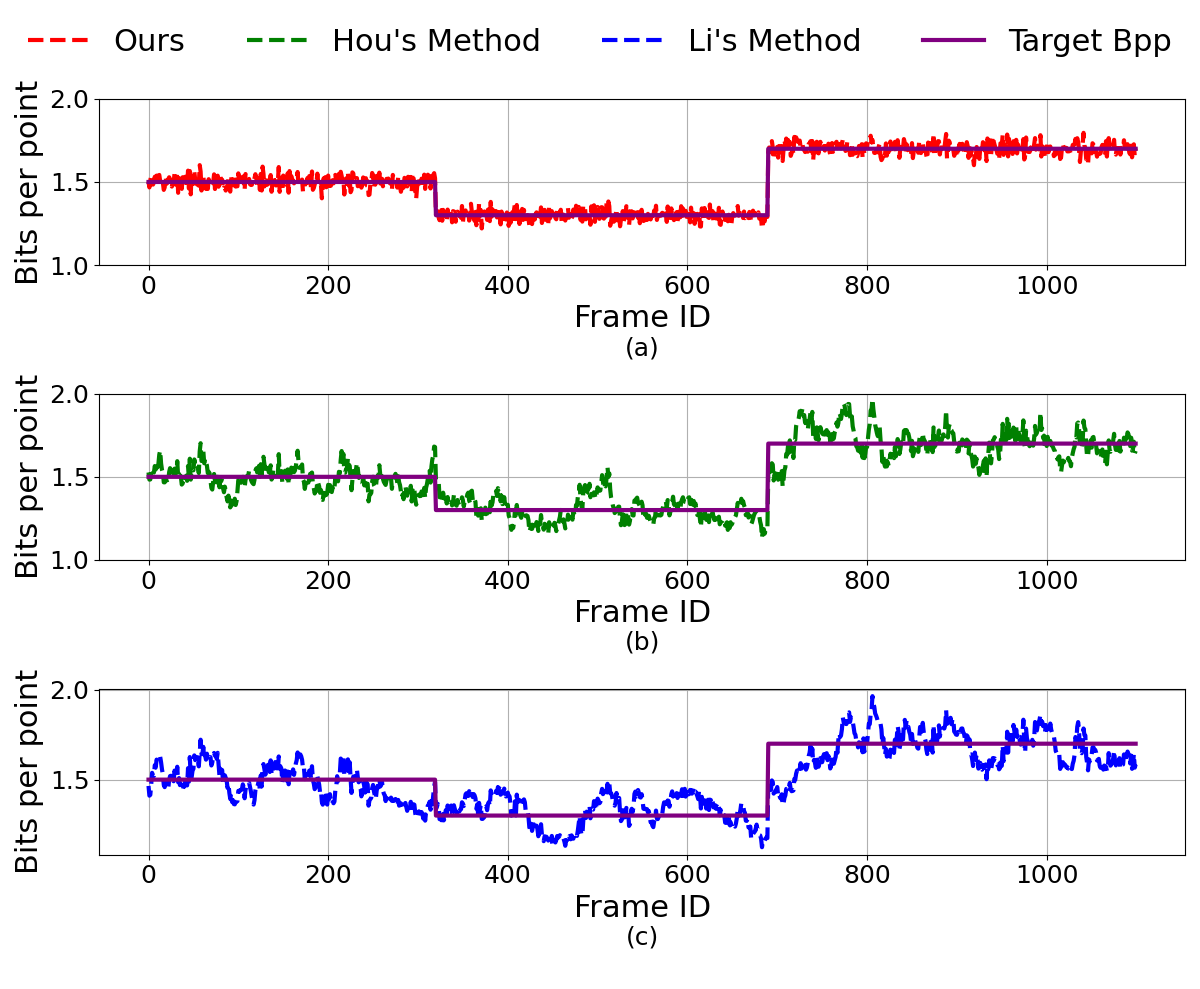}
    \caption{Rate control performance on $\mathrm{KITTI_{06}}$.}
    \label{fig: rate control}
\end{figure}

Fig.~\ref{fig: rate control} depicts the rate control performance for the $\mathrm{KITTI_{06}}$ sequence, where D-Compress exhibits greater bitrate adherence. Table.~\ref{table: rate control} summarizes the average and peak BEs across different dataset sequences. Our system achieves lowest average BEs of 2.9$\%$, 2.4$\%$, and 3.2$\%$ on the KITTI, NuScenes, and Waymo datasets, respectively. In contrast, Hou's method and Li's method exhibit significantly higher average errors of 5.3$\%$/6.2$\%$/6.9$\%$ and 9.8$\%$/8.5$\%$/9.7$\%$, respectively. More remarkably, D-Compress achieves a substantial reduction in peak BEs, maintaining levels as low as 9.2$\%$, 7.8$\%$, and 9.5$\%$. Conversely, Hou's method and Li's method suffer from considerably higher peak errors of 30.2$\%$/29.9$\%$/30.3$\%$ and 38.6$\%$/34.5$\%$/45.2$\%$, respectively.

\begin{table}
    \centering
    \caption{Bit error across different dataset sequences}
        \begin{tabular}{c|ccc}
            \hline
            \multirow{2}{*}{Dataset} & \multicolumn{3}{|c}{Average Bit Error/Peak Bit Error} \\
            & D-Compress & \makecell{Hou's \\ Method~\cite{hou2024rate}} & \makecell{Li's \\ Method~\cite{li2014lambda}}\\
            \hline
            KITTI$_\mathrm{03}$ & 2.9$\%$/8.4$\%$ & 5.5$\%$/29.4$\%$ & 8.7$\%$/29.9$\%$  \\
            KITTI$_\mathrm{06}$ & 1.7$\%$/6.8$\%$ & 4.3$\%$/19.9$\%$ & 5.6$\%$/20.0$\%$  \\
            KITTI$_\mathrm{09}$ & 4.4$\%$/11.6$\%$ & 5.9$\%$/43.3$\%$ & 14.6$\%$/53.3$\%$ \\
            KITTI$_\mathrm{12}$ & 2.5$\%$/8.7$\%$ & 5.0$\%$/32.9$\%$ & 12.3$\%$/44.8$\%$  \\
            KITTI$_\mathrm{15}$ & 3.3$\%$/10.2$\%$ & 5.8$\%$/31.5$\%$ & 9.0$\%$/47.3$\%$ \\
            KITTI$_\mathrm{18}$ & 2.4$\%$/9.3$\%$ & 5.4$\%$/24.2$\%$ & 8.6$\%$/36.2$\%$\\
            \textbf{Average} & \textbf{2.9$\%$/9.2$\%$} & \textbf{5.3$\%$/30.2$\%$} & \textbf{9.8$\%$/38.6$\%$} \\
            \hline
            Scenes$_\mathrm{mini}$ & 2.3$\%$/7.4$\%$ & 5.8$\%$/27.1$\%$ & 8.9$\%$/36.8$\%$ \\
            nuScenes$_\mathrm{06}$ & 2.5$\%$/8.3$\%$ & 6.6$\%$/32.7$\%$ & 8.1$\%$/32.2$\%$ \\
            \textbf{Average} & \textbf{2.4$\%$/7.8$\%$} & \textbf{6.2$\%$/29.9$\%$} & \textbf{8.5$\%$/34.5$\%$} \\
            \hline
            Waymo & 3.2$\%$/9.5$\%$ & 6.9$\%$/30.3$\%$ & 9.7$\%$/45.2$\%$ \\
            \hline
        \end{tabular}
    \label{table: rate control}
    \vspace{-1.5em}
\end{table}

\subsection{Ablation Study}
To validate the effectiveness of the a-DWT module, we solely replaced it with standard DWT and DCT algorithms commonly used in image/video codecs and compared the PSNR across different bpps using the KITTI dataset. As shown in Table.~\ref{table: ablation study}, the proposed a-DWT module significantly improves PSNR across all tested bpps by better preserving high-frequency details in range images, leading to enhanced compression accuracy. Simultaneously, our adaptive quantization process maintains computational latency comparable to that of DWT and DCT, without introducing significant computational overhead.
\begin{table}[h]
    \centering
    \caption{Comparison of PSNR  of a-DWT, DWT and DCT across various bpp.}
        \begin{tabular}{c|c@{\hskip 5pt}c@{\hskip 5pt}c@{\hskip 5pt}c@{\hskip 5pt}c|c}
            \hline
            \multirow{2}{*}{Method} & \multicolumn{5}{c|}{PSNR (dB)} & \multirow{2}{*}{\makecell{Runtime \\ (ms)}} \\
            \cline{2-6}
            & 0.6 bpp & 1.0 bpp & 1.4 bpp & 1.8 bpp & 2.2 bpp & \\
            \hline
            a-DWT & 57.9 & 63.1 & 67.3 & 69.6 & 72.4 & 3.6 \\
            DWT & 47.5 & 54.3 & 57.5 & 61.3 & 65.8 & 3.5 \\
            DCT & 34.7 & 41.8 & 46.8 & 51.4 & 55.9 & 2.4 \\
            \hline
        \end{tabular}
    \label{table: ablation study}
    \vspace{-1.5em}
\end{table}

\section{CONCLUSIONS and FUTURE WORK}
This paper presents D-Compress, a detail-preserving, real-time compression method for LiDAR range images. D-Compress combines intra- and inter-frame compression, and devises a precise adaptive discrete wavelet transform (a-DWT) method for residual image compression. Extensive experiments demonstrate that D-Compress outperforms state-of-the-art LiDAR point cloud compression (LPCC) methods in both accuracy and application-level performance, particularly under high compression ratios. Additionally, we propose a new rate control algorithm based on rate-distortion (R-D) modeling and optimization, enabling stable performance under dynamic bandwidth constraints. Real-time validation on a resource-constrained mini-PC shows that D-Compress achieves over 10 FPS without IMU data and exceeds 20 FPS with IMU assistance. In future work, we will explore the potential of hardware acceleration to further enhance real-time performance and address challenging scenarios, such as rapid robot rotations and highly dynamic environments.

\bibliographystyle{IEEEtran}
\bibliography{reference}

\end{document}